\documentclass[usenatbib]{mn2e}
\usepackage{rotating}
\usepackage{txfonts}
\usepackage{graphicx}
\usepackage{multirow}

\def\cmM2{\rm cm$^{-2}$}

\def\4c{4C04.42}
 
\def\chandra{{\it Chandra}}
\def\integral{{\it INTEGRAL}}
\def\xmm{{\it XMM-Newton}}

\def\asca{{\it ASCA}}
\def\swift{{\it Swift}}

\title[Bulk Compton motion in the luminous quasar 4C04.42?]
{Bulk Compton motion in the luminous quasar 4C04.42?}
\author[A. De Rosa et al.]{A. De Rosa$^{1}$\thanks{E-mail: alessandra.derosa@iasf-roma.inaf.it}, L. Bassani$^{2}$, P. Ubertini$^{1}$, A. Malizia$^{2}$,  A. J. Dean${^3}$\\
$^{1}$INAF/IASF-Roma, via del Fosso del Cavaliere 100, I-00133 Rome, Italy \\
$^{2}$INAF/IASF-Bologna, via Gobetti 101,I-40129 Bologna, Italy\\
$^{3}$School of Physics and Astronomy, University of Southampton
Highfield, Southampton, United Kingdom}

\begin{document}

\pagerange{\pageref{firstpage}--\pageref{lastpage}} \pubyear{2002}

\maketitle

\label{firstpage}

\begin{abstract}
We present the broadband analysis of the powerful quasar \4c (z=0.965)  observed by \xmm\, and \integral. The 0.2--200 keV spectrum is well reproduced with a hard power-law component ($\Gamma\sim$1.2), augmented by a soft component below 2 keV (observer frame), which is described by a thermal blackbody with temperature kT$\backsimeq$ 0.15 keV. Alternatively, a broken power-law with E$_{break}$=2 keV and $\Delta\Gamma$=0.4 can equally well describe the data.
Using archival data we compile the not-simultaneous  Spectral Energy Distribution of the source from radio to gamma-ray frequencies. The SED shows two main components: the low frequency one produced by Synchrotron radiation from the electrons moving in the jet and the high energy one produced through external Compton scattering of the electrons with the photon field of the Broad Line Region. Within this scenario the excess emission in the soft-X ray band can be interpreted as due to Bulk Compton radiation of cold electrons. However, some other processes, briefly discussed in the text, can also reproduce the observed bump.

\end{abstract}


\begin{table*}
\begin{flushleft}
\caption{Jurnal of the \xmm\, and \integral\, observation of \4c}
\begin{tabular}{lcccc}
\noalign{\hrule}
Instrument & Date & Exp Time & RA, DEC (J2000) & Counts/s\\
            &        &  (s)  & (h m s, $^{\circ}$'") & (s$^{-1}$)\\ \hline
\hline 
\xmm-MOS1 (thin filter) & 2006/07/12 &  11515 &  12 22 22.50,+04 13 16.0 & $^\star$0.173$\pm$0.005\\
\xmm-MOS2 (thin filter) & 2006/07/12 &  11527  &  12 22 22.50,+04 13 16.0 & $^\star$0.213$\pm$0.006 \\
\xmm-pn (thin filter) & 2006/07/12 & 8842   &  12 22 22.50,+04 13 16.0 & $^\dagger$0.83$\pm$0.01 \\
\integral-isgri & -  &  100000 &  12 22 26.88, +04 15 21.6 & $^{\circ}$0.12$\pm$0.02 \\
\hline
\end{tabular}
\label{jurnal}

\small{$^\star$ In 0.5-10 keV. $^{\dagger}$ In 0.3-12 keV. In $^{\circ}$ In 20-100 keV for IBIS/ISGRI.}
\end{flushleft}
\end{table*}
\vspace{-0.6cm}
\section{Introduction}
\label{intro}

Blazars are the most powerful objects in the observable Universe, and are capable 
to emit from radio frequencies up to the extreme gamma-ray.
In the X-ray energy range, harder spectra are associated with the highest luminosity objects \citep{fossati98}, and Flat Spectrum Radio Quasars (FSRQ) are the most luminous class of Blazars.
The Spectral Energy Distribution (SED) of FSRQ exhibits two main peaks (one between 
the IR and soft X-ray frequencies and the other in the gamma-ray regime),
disclosing the presence of two main components: it is widely believed that the low energy one is due to 
the Synchrotron radiation of relativistic electrons in a jet, while the high 
energy one is due to Inverse Compton scattering (IC) of the same electrons 
with a photon field \citep{ghisellini98}. 
It is also believed that in FSRQ the IC is due to Compton scattering of photons 
external to the jet (external Compton radiation EC), probably produced in 
the accretion disk and reprocessed by the Broad Line Region (BLR) and/or the dusty torus \citep{dermer93,wagner95}. 
Other competitive processes, like the IC from photons produced by Synchrotron 
(Synchrotron Self-Compton radiation, SSC) can contribute to the high 
energy component \citep{maraschi92}.
The investigation of these extreme objects in the X-ray band
is important  mainly because they emit most of their bolometric 
luminosity in this energy range 
\citep{elvis94} and because the fast variability  indicates 
that this radiation is produced very close to the central black-hole. 
Observations have demonstrated that radio-loud QSOs are more powerful  X-ray 
emitters than radio-quiet counterparts or RQQs \citep{page05} 
and, therefore, more easily observable.
Various authors \citep{reeves_turner00,piconcelli05} have demonstrated  
that radio-loud QSOs exhibit a flatter intrinsic spectral shape 
($\Gamma\sim$1.6) than RQQs ($\Gamma\sim$1.9); this harder 
X-ray emission is thought to be Synchrotron or IC emission 
from the relativistic radio jet, rather than from the accretion disk.

A deficit of photons in the soft X-ray band (E$<$ 2 keV observer frame) of several 
high and low-z QSOs  has been observed in the past by \asca\, \citep{fabian98,fiore98}, 
and recently confirmed through \xmm, \chandra\ and \swift\
observations \citep{bassani07,yuan06,sambruna07}.
The origin of this feature is not yet clear. Possible hypotheses are an intrinsic 
cold/ionized absorption or an intrinsic break of the continuum. When associated with
absorption a clear trend of N$_{\rm H}$ vs z has been measured, indicating a 
cosmic evolution effect, which seems to be strongest at redshifts 
around 2 \citep{yuan06}.
However, excess emission at similar low energies has also been observed, but only in a few sources \citep{kataoka07,sambruna06}. 
The origin of this behaviour is still unclear, even if several scenarios 
have been proposed and explored, like Bulk Comptonization \citep{celotti07}, increasing contribution of the SSC component, \citep{kataoka07}, and reflection from the accretion disk \citep{sambruna06}. 
So far these two spectral features (excess or deficit of soft X-ray  photons), have been interpreted as due to different emission processes. However, very recentely, \citet{celotti07} modelled the soft X-ray flattening observed in the quasar BG B1428-4217, via bulk Comptonization process.
Clearly more sources need to be observed in order to improve the 
statistics and study their physical properties in more detail.

\4c is a FSRQ, with radio loudness parameter 
$R$=S$_{5GHz}$/S$_{4400\dot{A}}\sim$ 1000 \citep{kellermann89}.
It was detected above 20 keV by \integral\ \citep{bird07} with a flux of
(7.6$\pm$1.5)$\times 10^{-12}$ erg cm$^{-2}$ s$^{-1}$ and 
(18.8$\pm$2.8)$\times 10^{-12}$ erg cm$^{-2}$ s$^{-1}$ in 20-40 keV 
and 40-100 keV bands respectively. It was later observed by \xmm\ 
for $\sim$10 ks.
In this letter we present  the broadband analysis of \4c 
in 0.2-200 keV, for the first time, through non simultaneous  \xmm\ and \integral\ data
and provide strong evidence for soft excess emission below 2 keV.
In Sect. \ref{obs} we present the observations and data reduction, in Sect. 
\ref{analysis} we describe the detailed spectral analysis, 
while the discussion is reported in Sect. \ref{discussion}. 
We draw our conclusions in Sect. \ref{conclusions}.  
Throughout this paper we adopt a luminosity distance of 
d$_{L}$ = 6358 Mpc for \4c (z = 0.965), derived for a $\Lambda$CDM 
cosmology with $\Omega_{m}$ = 0.27, $\Omega_{\Lambda}$= 0.73 
and H$_{0}$ = 71 km s$^{-1}$ Mpc$^{-1}$.
\vspace{-0.6cm}
\section{Observations and data reduction}
\label{obs}
\subsection{\xmm}
\xmm\, observed \4c in July 2006 and details of this observation are given in 
Table \ref{jurnal}. EPIC data reduction was performed with version 7.0 
of the SAS software, employing the most updated calibration files available 
at the time of the data reduction (2007 May).
Patterns 0 to 12 (4) were employed in the extraction of the MOS (pn)
scientific products. EPIC-pn and EPIC-MOS spectra for the source and background were
extracted from a circular region of 30 arcsec radius around the source and 
from a source-free region respectively.
Spectra were rebinned to have at least 20 counts in each spectral
channel.
Spectral fits were performed simultaneously with MOS1, MOS2 (0.3--9 keV) 
and pn (0.8-10 keV).
The results obtained fitting the data of MOS1/MOS2 and pn separately
will also be discussed in the Sect. \ref{analysis}.
\vspace{-0.8cm}
\subsection{\integral}

\emph{INTEGRAL}-IBIS \citep{ube03} data have been reduced following the same 
procedure used for the 
survey work and described in \citet{bird07}. The source was detected with a 
significance of $\sim$ 8$\sigma$.
First IBIS/ISGRI images for each available pointing were generated in 13 energy
bands using the ISDC offline scientific analysis software OSA
version 5.1. Count rates at the source position were extracted
from individual images in order to provide light curves
in the various energy bands sampled. Since the light curves did not show any 
sign of variability or flaring activity, average fluxes were
then extracted in each band and combined to produce the source spectrum. 
We tested the reliability of this spectral extraction method by comparing the Crab spectrum obtained in this way with the one extracted
using the standard spectral analysis.

\begin{figure}
\centering
\includegraphics[width=0.65\linewidth, angle=-90]{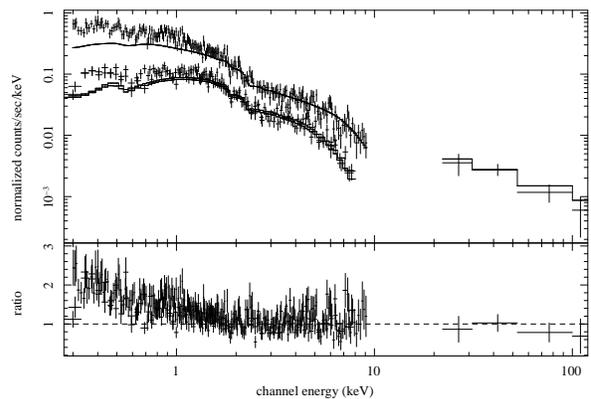}
\caption{Continuum power-law fit to the energy band above 2 keV (observer frame) extrapolated  over the low energy range. \textit{Top curve}: pn data, \textit{bottom curve}: MOS.}
\label{data ratio}
\end{figure}
%
\begin{table*}
\begin{flushleft}
\caption{Best-fit parameters of the different models we used to reproduce the combined broadband (0.2--200 keV) \xmm\ and \integral\, data.}
\begin{tabular}{cccccccc}
\noalign{\hrule}
$^{1}$C$_{integral}$ & $\chi^2$/dof  &  $^2$P$_{null}$ & $\Gamma$ & F$^{obs}$(0.1-2 keV) &
F$^{obs}$(2-10 keV) & F$^{obs}$(20-100 keV) & param \\ 
& & & & (\rm 10$^{-12}$ erg cm$^{-2}$ s$^{-1}$) & (\rm 10$^{-12}$ erg cm$^{-2}$ s$^{-1}$) & (\rm 10$^{-12}$ erg cm$^{-2}$ s$^{-1}$)&  \\
\hline\hline
\multicolumn{7}{c}{constant phabs zpowerlw} &\\ 
\hline\hline
 1 frozen & 428/366 & 0.01 & $1.44^{+0.03}_{-0.03}$ & 1.0 & 2.2 & 8.2 & $-$ \\
 $1.9\pm0.5$ & 419/365 & 0.03 & $1.44^{+0.03}_{-0.03}$ & 1.0 & 2.2 & 8.0 & $-$ \\
\hline\hline
\multicolumn{7}{c}{constant phabs zpowerlw zbb} & $^3$kT$_{BB}$\\ 
\hline\hline
 1 frozen & 383/364& 0.232 & $1.28^{+0.02}_{-0.04}$ & 0.98 & 2.5 & 13 & $0.15^{+0.02}_{-0.02}$ \\
 $1.2^{+0.3}_{-0.3}$ & 383/363 & 0.229 & $1.30^{+0.04}_{-0.03}$ & 0.99 & 2.5 & 13.  & $0.15^{+0.02}_{-0.02}$ \\
\hline\hline
\multicolumn{7}{c}{constant phabs zpowerlw diskbb} & $^3$kT$_{inner}$\\ 
\hline\hline
 1 frozen & 373/364 & 0.363 & $1.22^{+0.06}_{-0.06}$ & 1.0 & 2.5 & 15 & $0.25^{+0.04}_{-0.04}$ \\
 $0.97^{+0.04}_{-0.03}$ & 373/363 & 0.349 & $1.22^{+0.07}_{-0.08}$ & 1 & 2.5 & 15 & $0.26^{+0.04}_{-0.04}$ \\
\hline\hline
\multicolumn{7}{c}{constant phabs brokenpowerlw} & $^3$E$_{bkn}$.\\ 
\hline\hline
 1 frozen & 365/364 & 0.476 & $1.58^{+0.05}_{-0.04}$ & 1.0 & 2.5 & 16.  & $2.1^{+0.4}_{-0.3}$ \\
    &  &  & $1.19^{+0.07}_{-0.05}$ &  &  &  &  \\
 $0.9^{+0.3}_{-0.3}$ & 364/363 & 0.470 & $1.59^{+0.05}_{-0.05}$ & 1.0 & 2.6 & 17 & $2.2^{+0.4}_{-0.4}$  \\
    &  &  & $1.17^{+0.09}_{-0.10}$ &  &  &   &  \\
\hline\hline
\multicolumn{7}{c}{constant phabs  zbb pexrav} & R\\ 
\hline\hline
 1 frozen & 382/363 & 0.233 & $1.33^{+0.09}_{-0.08}$ & 1.0 & 2.6 & 14 & $0.25^{+0.42}_{-0.24}$ \\
 $1.1^{+0.4}_{-0.3}$ & 382/362 & 0.224 & $1.33^{+0.07}_{-0.08}$ & 1.0 & 2.5 & 14 & $<$0.6 \\
\hline
\hline
\end{tabular}
\label{fit}

\small{
$^1$ Cross-calibration constant \integral-isgri/\xmm-MOS1; 
$ ^2$ Null hypothesis probability.
$ ^3$ in keV.}
\end{flushleft}
\end{table*}
\vspace{-0.6cm}
\section{\xmm\, and \integral\, data analysis}
\label{analysis}
We performed spectral analysis with \texttt{XSPEC v11.3}. Errors
correspond to 90 per cent confidence level for one interesting parameter
($\Delta\chi^2$=2.7). All parameters values in the text and tables refer to those measured in 
the observer frame, if not otherwise specified.
To reproduce the intrinsic continuum of \4c we fitted \xmm\, and \integral\, data over the 2-150 keV band with an absorbed power-law 
and obtain a good fit with photon index $\Gamma$=1.21$^{+0.06}_{-0.06}$ 
and the column density fixed to the Galactic value (N$_{\rm H}$=1.69$\times$10$^{20}$ \cmM2). 
This absorbed power-law, extrapolated to the 0.3--2 keV band, fails to reproduce the 
observed spectrum (see Figure  \ref{data ratio}, and first line in Table \ref{fit}), 
giving a $\chi^{2}$/dof=428/366. 
Major residuals are seen at low energies, strongly suggesting the 
presence of excess emission below 2 keV. To fit more accurately the  combined  
\xmm\ and \integral\ spectrum, and to fully account for the extra low energies component, 
we tried a range of different models.
In Table \ref{fit} we report the best fit parameters for each of them. 
Fitting the excess with a thermal black-body model in addition to the 
absorbed power-law we obtain a good result  ($\chi^2/dof$=383/364),
with kT$_{BB}$ around 0.15 keV.
A multiple blackbody component (\texttt{diskbb} in \texttt{XSPEC}) reproduces the excess even better that the single temperature black-body 
($\chi^2/dof$=373/364).
We  also tested the possibility that the observed behaviour is due to a 
break in the 
intrinsic continuum. We substituted the primary power-law with a 
broken power-law 
without including any other emission component at low energies. Following this representation we found that, at energies below E$_{break}\sim$2 keV, the continuum photon index is steeper  than above E$_{break}$, with $\Delta\Gamma\sim$ 0.4 (see Table \ref{fit}).
In this case we find a good fit, with $\Delta\chi^{2}$=8 with respect to 
the multiple blackbody model. However, the null hypothesis probability in this case, increases  by only $\sim$10 per cent with respect to the \texttt{diskbb} model,  suggesting that  strong statistical evidence does not exist for broken power-law to be preferred over  the thermal one.
The luminosity of the black-body component in the 0.1-3 keV band is L$_{BBody}\sim$10$^{45}$ erg s$^{-1}$.
Testing for the presence of a Compton reflection component, using the model \texttt{pexrav} in 
\texttt{XSPEC} \citep{pexrav_ref}, we find only an upper limit for the relative reflection 
(R=$\Omega$/2$\pi<$0.6). No evidence for an iron line is found in the data. 
Properly redshifted, we included a narrow gaussian line at the energy of cold Fe, 
obtaining an upper limit for the line intensity 
I$_{\rm Fe} \le$ 10$^{-6}$ ph cm$^{-2}$ s$^{-1}$ (i.e. EW$_{\rm Fe} \le$ 15 eV).
We stress here that the estimated value of the Compton
reflection fraction R is strongly dependent on the value of the cross-calibration constant  
C$_{integral}$ between the soft gamma (\integral) and X-ray  (\xmm) data.
In this respect, we note that Kirsch  and collaborators (2005) analysed the 
Crab spectrum observed with different instruments and  concluded that 
at 20 keV the  \xmm/\integral\,  cross-calibration was close to 1 
(within few per cent). However we also checked this assumption a posteriori
repeating all the  spectral fits, leaving C$_{integral}$ free to vary. The best fit parameters 
are also included in Table \ref{fit}. The value of C$_{integral}$  
is always well constrained around one and the main results of the analysis we have presented 
do not change leaving C$_{integral}$ free to vary.
The same outcome is found if the combined MOS1/MOS2 
and pn data are fitted separately  with that of \integral. 
The only very marginal effect is that the MOS1/MOS2-\integral\, spectrum provides 
(in all spectral models presented) a flatter photon index than  the one obtained with the 
pn-\integral\, spectrum. The flattening of $\Gamma$ is however less than 10\%.
Similarly, the values of the spectral parameters in Table \ref{fit}, do not change if we use only the \xmm\, data. Furthermore, even the range of uncertainty of these parameters does not change significantly. In fact, the statistics in the broadband analysis is fully dominated by the higher quality of the \xmm\, data.
\vspace{-0.6cm}
\section{Discussion}
\label{discussion}
\begin{figure}
\centering
\includegraphics[width=0.8\linewidth]{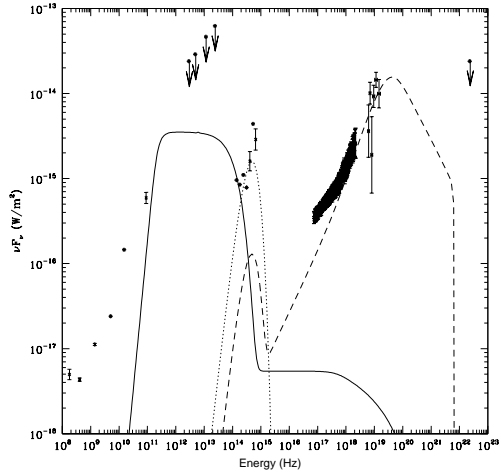}
\caption{Spectral Energy Distribution from radio up to soft gamma-ray frequencies. Data are from literature taken by NED}
\label{SED}
\end{figure}

In Figure \ref{SED} we show the SED of \4c where non simultaneous data across the electromagnetic spectrum 
have been taken from NED\footnote{NRAO Extragalactic Database; http://nedwww.ipac.caltech.edu/index.html}.
In FSRQ the two peaked SED is interpreted in the Synchrotron External Compton 
framework \citep{ghisellini98}: the low frequency peak in the IR or FIR is created through 
the Synchrotron radiation of relativistic electrons moving in a blob, of dimension R, 
originating in the jet and a magnetic field of few Gauss. The high 
frequency peak in the hard-X  and gamma-rays is believed to be the effect of IC 
scattering of the same electrons population on a photon field that can be 
caused (1) by the Synchrotron radiation (SSC) and/or 
(2) by thermal emission from the accretion disk, BLR and/or dusty torus (EC). 
Due to the high density of  the external photon field, the EC in FSRQ dominates 
over  the SSC.  
The secondary effect on the SED derived from 
EC in the molecular torus, distant on the scale of pc from the central black hole, is not considered here.
A model that well  reproduce the SED\footnote{SSC code from http://www.asdc.asi.it/ssc$_{-}$at/. The model was used in e.g. \citet{massaro06} and \citet{derosa05}} is also included in Figure \ref{SED}. 
The simplest hypothesis is that the emitting region is a blob of radius R 
moving with a bulk velocity of $\beta$c 
(where $\beta=\sqrt{\Gamma^{2}-1}/\Gamma$, and $\Gamma$ is the Lorentz factor),
in a magnetic field with intensity of 2 Gauss. We assume an observing angle of $\theta\sim1/\Gamma$ implying a Doppler factor $\delta=\Gamma$.
The SSC+EC components are plotted separately in Figure \ref{SED} with solid 
and dashed lines respectively. The disk contribution (dotted line) is shown
as a thermal component with temperature of kT$_{BB,z}=$1 eV in the quasar 
rest frame. No other additional component, such as reflection from the disk, has been 
included in the model. 
The dimension of the emitting blob, the magnetic field, the electron energy 
distribution, and the Lorentz factor are the only parameters required to evaluate the 
SSC component in the SED. We assume that the electrons in the blob  are well described by
a power-law, $N(\gamma)$=N $\gamma^{-p}$ cm$^{-3}$, in the range $\gamma_{min}$ 
and $\gamma_{max}$, and we do not take into account radiative cooling of the electrons. The input parameters for our model are reported in 
the Table \ref{SED model}.
We stress that our model for the whole SED is not a true ''fit''. In view of the not simultaneous data set,  we just attempt to extract some physical parameters from the data, with the main goal to obtain average properties of the source. In particular the EC component did allow us to constrain  $\gamma_{max}$,  and  p.
The EC component due to the BLR photon field is determined 
by the BLR energy density U$_{BLR}$. We fixed the disk luminosity  at the 
level of the optical-UV emission (at 10$^{15}$ Hz in the quasar rest frame)
observed in the source SED. The BLR luminosity is estimated 
assuming that that region reprocesses a fraction of the disk luminosity that 
is $\sim$0.1 \citep{maraschi_tavecchio_03}. The energy density in the observer 
frame is given by  U$^{oss}_{BLR}$=L$_{BLR}$/4$\pi$cR$_{BLR}^2$, so that
the contribution of the IC emission to the SED is  
P$_{EC}$=4/3$\sigma_{T}$c$\delta^{4}\int_{\gamma_{min}}^{\gamma_{max}} 
N(\gamma) \gamma^{2} U^{oss}_{BLR} \Gamma^{2} d\gamma$= 8$\times$10$^{-2}$ R$_{BLR,17}^{-2}$ 
erg cm$^{-3}$ s$^{-1}$, where R$_{BLR,17}$ is the BLR extension in units 
of 10$^{17}$ cm. R$_{BLR}$  is measured in very few AGN through 
reverberation mapping \citep{kaspi05}, where
 a relation between R$_{BLR}$ and $\lambda$L$_\lambda$(5100 $\dot{A}$) is assumed.
Taking $\lambda$L$_\lambda$(5100 $\dot{A}$) from the SED, we have R$_{BLR}$=5$\times$10$^{17}$ cm. In the emitting 
volume given by V=4/3$\pi$R$^{3}$, we would then expect L$_{EC}$=3$\times$10$^{47}$ erg s$^{-1}$, in agreement with the value reported in the SED.

Figure \ref{SED}  clearly shows evidence that the
SSC+EC  model does not well fit the whole SED. In particular  an excess, of about of a factor of three with respect to the EC component, is visible around 2 keV. 
This soft X-ray excess could be due to bulk Compton of  a "cold" shell of plasma moving in the jet and interacting with the external BLR photon field \citep{celotti07}.  
The luminosities of the Bulk and External Compton components, L$_{BC}$  and L$_{EC}$, are determined by the relative normalizations of the number densities of the cold and relativistic electrons. 
The number of cold electrons is usually difficult to verify since low 
energy electrons emit synchrotron light
in the self-absorbed range, and SSC radiation at  low frequencies, where the flux is dominated
by the Synchrotron emission of electrons of higher energies. 
The only way to observe the bulk Compton component in the SED is when it dominates the non thermal continuum in the soft X-ray energy range, and this will happen when the competitive processes, like SSC, are negligible in this band. 
However, here we do not try to fit this excess with a bulk Compton component, we just stress that
with $\Gamma$=20, as assumed in our model, we expect to detect the peak of such component at the frequency of $\nu_{BC}=\Gamma^{2}\nu_{BLR}/(1+z)\sim$ 1 keV, i.e. where the excess is observed. In addition, if the excess is interpreted as due to bulk motion then L$_{BC}$=4/3$\sigma_{T}$c$\delta^{4}$ U$^{oss}_{BLR} \Gamma^{2}$N$_{c}$ $\sim$ 10$^{46}$N$_{c}$/N, where N$_{c}$ is the number of cold electrons in the blob. With the value obtained in  Sect. \ref{analysis}, L$_{BC}\sim$10$^{45}$ erg s$^{-1}$, we estimate the ratio N$_{c}$/N to be $\sim$ 0.1.
Very recently the bulk Compton process has been proposed as a means to explain the steepening of the spectrum in the case of quasars PKS~1510-089 at z=0.361 \citep{kataoka07}, 0723-679 at z=0.847, 1136-135 at z=0.554, and 1150-497  at z=0.334 \citep{sambruna06}.
The bulk Compton process has been  also proposed to explain the observed flattening in the powerful blazar BG B1428+4217, at z=4.72 \citep{celotti07}. In this source the soft X-ray behaviour was previously  attributed to warm absorption intrinsic to the source \citep{fabian98,fabian01,yuan06}. 
We also stress here that flattening of the spectral emission has been observed in high-z quasars up to z=4.4, while the softening has been observed in very few sources, and all are at redshifts below 1 (see Table \ref{sample}).
\begin{table}
\caption{SED model parameters for \4c}
\begin{tabular}{cccccccc}
\noalign{\hrule}
$^{1}$B & $^{2}$R & $^{3}\delta$ & $^{4}$N &  $^{5}\gamma_{min}$/$\gamma_{max}$ & $^{6}$p & $^{7}$L$_{BLR}$ & $^{8}$R$_{BLR}$\\
(Gauss) & (cm) & (=$\Gamma$) & (cm$^{-3}$) &  & & (erg s$^{-1}$) & (10$^{17}$ cm) \\
\hline\hline
2 & 3$\times$10$^{16}$ & 20 & 10$^{4}$& 1/1000 & 3 & 2$\times$10$^{45}$ & 5\\
\hline\hline
\end{tabular}
\label{SED model}
\small{
$^1$Magnetic field; 
$ ^2$Dimension of the emitting blob;
$ ^3$Doppler factor assumed to be equal to the bulk Lorentz factor;
$ ^4$number density of the electrons; $ ^5$minimum/maximum electron Lorentz factor; 
$ ^6$electron spectral index; $ ^7$ observed BLR Luminosity; $ ^8$BLR extension.}
\end{table}

Alternative models to bulk Comptonization  can be taken into account:
(a) an increasing contribution of the SSC component in the energy 
band between the Synchrotron and EC peaks, as already proposed by \citet{kataoka07}. However the SED associated with the model in Figure \ref{SED} makes  this hypothesis unlikely in the case of \4c;
(b) the emission originating in the central region of the accretion flow 
\citep{done_naya07}, as observed in several non-blazar objects;
(c) the Synchrotron peak placed in the in the X-ray band and not at UV frequencies, as observed in some FSRQ  \citep{padovani03, giommi07}. However, this does not seem to be the case for \4c, in fact,  the SED indicates that this peak is in the IR range; (d) the disk emission, assuming that it was composed by a black body from optical to soft X-ray (like in b), plus a flat spectrum reproducing the reflection component at harder X-ray, as proposed by \citet{sambruna06} to resolve the excess of soft X-ray emission  in three powerful radio loud quasars, 1136-135, 1150-497, 0723-679; 
(e) \citet{tavecchio_ghisellini08}, 
have recently demonstrated that the EC radiation, computed taking into account a realistic spectrum
for the external radiation originated in the BLR (calculated with the 
photoionization code \texttt{CLOUDY}, \citet{ferland98}), shows evidence of a 
steepening below 1 keV, that does not appear when the EC radiation is computed using 
a simple black body parameterization for the BLR emission.

Important physical quantities can be derived by a comparison between the power carried by the jet P$_{jet}$, and the disk luminosity L$_{disk}$.
Under the assumption of one proton per emitting electron in the jet 
composition we can estimate  the jet kinetic power 
P$_{jet}$=$\pi$R$^2\beta\Gamma^2$cU for \4c, where U=U$_B$+U$_e$+U$_p$
is the total energy in the jet rest frame due to magnetic field, 
electron and protons: 
U$_e$+U$_p$=n$_e$m$_e$c$^2 [\langle\gamma\rangle+n_p/n_e(m_p/m_e)]$, 
with n$_e=\int_{\gamma_{min}}^{\gamma_{max}}N(\gamma)d\gamma$ and 
$\langle\gamma\rangle$ the average Lorentz factor of the electrons. In Table \ref{sample} we provide the  values of L$_{disk}$ and P$_{jet}$ for ''similar'' sources, i.e. those showing an excess at low energies.
For \4c, using the model we employed to reproduce the SED and $\gamma_{min}$=1, we obtain L$_{disk}$/P$_{jet}\sim$0.1 which is in very good agreement 
with the other cases in where this ratio was evaluated with good accuracy 
\citep{maraschi_tavecchio_03,sambruna06}.  In the plane L$_{disk}$ $vs$ P$_{jet}$, \4c lies in the region of high-luminosity blazars, following the trend observed in other similar sources. This, in turn, implies that a large fraction of the accretion power in converted in bulk kinetic energy of the jet \citep{maraschi_tavecchio_03,celotti_ghisellini08}.  We stress that increasing $\gamma_{min}$ to 10 P$_{jet}$ is augmented by a factor of 10, leaving unchanged our findings.

Finally, we note that the combined broadband \xmm\, and \integral\, spectral analysis gives us a harder spectral index than typically observed in the radio loud QSOs \citep{reeves_turner00,piconcelli05}. 
This evidence suggests that this source could be  a member
of the small subclass of ''MeV blazars'' \citep{bloemen95,tavecchio00}. Future observation in the gamma-ray band with GLAST and AGILE will be able to address this issue.

\begin{table}
\caption{Comparison between the sources that show the excess emission below 2 keV.}
\begin{tabular}{ccccc}
\noalign{\hrule}
Source & z& R$_{BLR}$ & L$_{disk}$ &  $^{\star}$P$_{jet}$ \\
 & & (10$^{17}$ cm) &  (10$^{45}$ erg s$^{-1}$) & (10$^{45}$ erg s$^{-1}$) \\
\hline 
\4c & 0.965 & 5 & 20 & 300\\
$^{(1)}$0723+679 & 0.847 & 2 &10 & 350 \\
$^{(1)}$1136-135 & 0.554 & 5.2 & 5 & 73 \\
$^{(1)}$1150-497 & 0.334 & 2.6 & 3 & 76 \\
$^{(2)}$1510-089 & 0.361 & 5  &4 & 500\\
\hline
\end{tabular}
\label{sample}
\small{$^{(1)}$ \citet{sambruna06}; $^{(2)}$ \cite{maraschi_tavecchio_03,kataoka07}. $^{\star}$ calculated for n$_{e}$=n$_{p}$. See text for details.}
\end{table}
\vspace{-0.4cm}
\section{Conclusions}
\label{conclusions}

We have presented the (non simultaneous) broadband spectral analysis of the powerful quasar \4c at z=0.965 observed by \xmm\ and \integral. Archival data from NED allowed us to build the Spectral Energy Distribution and to study possible emitting scenarios in the framework of the SSC+EC models.
Our main results are:

\begin{itemize}
\item
The 2--200 keV spectrum is best reproduced by a power-law model, with photon index $\sim$ 1.2,  i.e. flatter than observed in radio loud QSOs. 

\item
The spectrum below 2 keV (observer frame) clearly shows evidence for a steepening, and the broadband data  can be described either with a power--law plus a thermal component with kT$_{BB}$=0.2 keV or by a broken power--law with E$_{break}$=2 keV and $\Delta\Gamma$=0.4. Both models are equally statistically acceptable.

\item
The doubled peaked SED built with non simultaneous data from radio up to gamma-ray, has been interpreted within a scenario that invokes a blob of relativistic electrons emitting Synchrotron radiation moving in a magnetic field of a few Gauss, plus IC radiation in the photon field of the BLR.

\item
The excess found in the soft X-ray, as well as the flat spectral shape in hard X-ray, strongly suggests the presence of a population of cold electrons able to produce a bulk Compton  feature at $\sim$ 1 keV  through IC with the photon field of the BLR. 

\item
The ratio L$_{disk}$/P$_{jet}\sim$0.1 between the disk luminosity and the power carried by the jet,  implies that a large fraction of the accretion power is converted in bulk kinetic energy of the jet.
\end{itemize}
This program is funded by Italian Space Agency  grant  via contracts I/008/07/0 and I/023/05/0. We thank the referee for constructive suggestions. 

\end{document}